\documentclass[referee]{aa}
\def\be{\begin{equation}}
\def\en{\end{equation}}
\input{psfig.sty}

\begin{document}

\thesaurus{02.08.1, 03.13.4, 02.18.8, 02.01.2}

\title{General Relativistic Hydrodynamics \\ with
Special Relativistic Riemann Solvers}

\author{Jos\'e A. Pons \inst{1} \and Jos\'e A. Font \inst{2} \and
Jos\'e M$^{\underline{\mbox{a}}}$. Ib\'a\~{n}ez \inst{1} \and \\
Jos\'e M$^{\underline{\mbox{a}}}$. Mart\'{\i} \inst{1} \and 
Juan A. Miralles \inst{1}} 

\offprints{Jose A. Pons}
\mail{Jose.A.Pons@uv.es}

\institute{Departament d'Astronomia i Astrof\'{\i}sica, 
Universitat de Val\`{e}ncia, \\ 46100 Burjassot (Val\`{e}ncia), Spain  
\and
Max-Planck-Institut f\"ur Gravitationsphysik, Albert-Einstein-Institut \\  
Schlaatzweg 1, 14473 Potsdam, Germany}

\date{Received date; Accepted date}

\authorrunning{Jose A. Pons {\it et al.}}
\titlerunning{GRH with SRRS}

\maketitle

\begin{abstract}
We present a general and practical procedure 
to solve the general relativistic hydrodynamic equations by 
using any of the special relativistic Riemann solvers 
recently developed for
describing the evolution of special relativistic flows. 
Our proposal relies on a local change of 
coordinates in terms of which the spacetime metric 
is locally Minkowskian and permits accurate 
numerical calculations
of general relativistic hydrodynamics problems using the
numerical tools developed for the special relativistic case 
with negligible computational cost.
The feasibility of the method has been confirmed by a number of 
numerical experiments.
\end{abstract}

\keywords{Hydrodynamics, Methods:numerical, Relativity, Accretion}

\section{Introduction}

In the near future the first generation 
of Earth-based laser-interferometer 
detectors of gravitational waves 
will be operating (LIGO, VIRGO, GEO600, TAMA).
This perspective has stimulated researchers working in numerical 
relativistic astrophysics to develop robust codes for the 
simulation of the different astrophysical sources of 
gravitational radiation, such as, e.g., 
stellar core collapse, coalescing compact binaries 
or accretion onto compact objects.

Relativistic hydrodynamical codes experienced a substantial advance at the
beginning of the nineties (Mart\'{\i}, Ib\'a\~{n}ez, Miralles,
\cite{MIM91}) with the implementation
of high--resolution shock--capturing methods (HRSC) originally developed
in classical fluid dynamics. These methods make use of the hyperbolic and
conservative character of the equations and
display a number of interesting features and properties, as being
stable and conservative, converging to physical solutions and having high
accuracy in regions where the solution is smooth. They are all based on
the resolution of local Riemann problems at the interfaces of numerical cells
--following the seminal idea of Godunov (\cite{Go59}) -- ensuring a consistent 
treatment
of discontinuities (shocks). The first relativistic applications of these
techniques showed their capabilities in describing accurately complex flows,
with high Lorentz factors and strong shocks, superseding traditional methods
(Wilson \cite{Wi79}) which failed 
to describe ultrarelativistic flows (Norman and Winkler \cite{NW86}).
Up to now, the use of HRSC methods in relativity has been
mainly restricted to the field of special relativistic hydrodynamics (SRH)
in the simulation of collisions of heavy ions and, 
remarkably, extragalactic jets (Mart\'{\i}, M\"uller and  
Ib\'a\~{n}ez, \cite{MMI94}).
We refer the interested reader to the introductory section in
Banyuls {\it et al.} (\cite{BFIMM97}) for a recent review 
of the current
status of HRSC techniques in numerical relativistic hydrodynamics.

In recent years, the great success obtained in the first relativistic
applications has drawn the attention of specialists who started to
develop specific Riemann solvers for SRH. Nowadays, most of the reliable
HRSC methods developed during the last twenty years in classical 
hydrodynamics have their special--relativistic extension 
(see, e.g., Ib\'a\~{n}ez {\it et al.},
\cite{IFMM97}, for an updated list).
In the case of general relativistic hydrodynamics (GRH), the development of
numerical codes based on the resolution of local Riemann problems
is still in its infancy.
Only a small number of papers have considered the
extension of HRSC methods from SRH to GRH (see below).
In addition, recently, several formulations 
of Einstein Equations as a first-order 
hyperbolic system of balance laws have been derived 
(Friedrich \cite{Fri85}; Bona {\it et al.} \cite{Bon95};
Abrahams {\it et al.} \cite{Abr95}; Fritelli \& Reula \cite{Fri96}).
This opens a new 
strategy, in the field of Numerical Relativity,
permitting the use of HRSC schemes, specifically
designed for such hyperbolic systems, to solve the coupled
system of spacetime plus hydrodynamics (Bona {\it et al.} 
\cite{escorial}).

The basic idea behind this work is to obtain a general
procedure  that allows us to take advantage of the increasing 
number of special relativistic Riemann solvers (SRRS) 
developed recently, in order to generate numerical solutions 
describing the evolution of relativistic flows in strong 
gravitational fields (background or dynamical).
All the previous works done to extend HRSC methods to GRH
have used linearized Riemann solvers
(Mart\'{\i}, Ib\'a\~{n}ez, Miralles \cite{MIM91};
Eulderink and Mellema \cite{EM94}; Romero {\it et al.} \cite{Retal96};
Banyuls {\it et al.} \cite{BFIMM97}).
In this paper we describe a procedure to use any type of SRRS 
in general relativistic hydrodynamics, including the exact 
Riemann solver for 1D problems.
This procedure relies on a local change of 
coordinates at each numerical interface,
in terms of which the spacetime metric is locally flat,  
analogously to the approach followed in classical fluid 
dynamics, when using the solution of Riemann problems in 
general curvilinear coordinates. The numerical implementation
is simple, computationally inexpensive, and provides a useful
tool for the researchers currently working in SRH to enter 
the field of GRH.

The structure of this paper is the following: In \S 2 we summarize
the GRH equations in the $\{3+1\}$ 
formalism and the basic ideas of HRSC methods. In \S 3 the
formulation of Riemann problems in locally flat spacetimes
and the method to obtain the numerical fluxes
for the GRH equations from those obtained in SRH is explained.
In \S 4 we briefly describe the set of numerical tests 
and applications performed
to demonstrate the feasibility of the approach.
Finally, in \S 5
we summarize our results and foresee other applications of
our proposal.

%%%%%%%%%%%%%%%%%%%%%%%%%%%%%%%%%%%%%%%%%%%%%%%%%%%%%%%%%%%%%
\section{GRH equations in the {3+1} 
formalism and HRSC methods.}

Let $\cal M$ be a general spacetime, described by the four
dimensional metric tensor $\bf g$. The line element has the form
\begin{equation}
ds^{2} = -(\alpha^{2}-\beta_{i}\beta^{i}) dt^{2}+
2 \beta_{i} dx^{i} dt + \gamma_{ij} dx^{i}dx^{j}
\end{equation}

\noindent
Throughout the paper Greek (Latin) subscripts run from 0 to 3 (1 to 3)
and geometrized units are used ($G=c=1$). 
Let $\{\partial_t,\partial_i \}$ be the coordinate basis
and $\bf n$ the unit timelike vector field normal to the
spacelike hypersurfaces $\Sigma_t$ (t = const.) 

\begin{equation}
\partial_t = \alpha {\bf n} + \beta^i \partial_i
\end{equation}

We denote by $\bf J$ and $\bf T$ the density 
current and the energy--momentum tensor 
for a perfect fluid, respectively,  
\begin{equation}
{\bf J} = \rho {\bf u} 
\end{equation} 
\begin{equation}
{\bf T} = \rho h {\bf u} \otimes {\bf u} + p {\bf g} 
\end{equation}
with $\bf u$ the 
four-velocity of the fluid, $\rho$ the rest--mass density, 
$p$ the pressure and $h$ the specific enthalpy ($h = 1 + \epsilon + p/\rho$, 
where $\epsilon$ is the specific internal energy). 

The equations describing the 
evolution of a relativistic fluid are the {\it local conservation laws} of 
baryon number and energy--momentum and can be written, for observers
which are at rest in the slice $\Sigma_t$ ({\it Eulerian observers}), in 
terms of the divergence of the 5 vector fields 
$\{ {\bf J}$, ${\bf T}\cdot {\bf n}$, ${\bf T}\cdot \partial_1$, 
${\bf T}\cdot \partial_2$, ${\bf T}\cdot \partial_3 \}$ as,

\begin{equation}
\nabla \cdot {\bf A} = s,
\label{diva}
\end{equation}

\noindent
where ${\bf A}$ denotes any of the above 5 vector fields and $s$ is the 
corresponding source term. Explicit expressions of these vectors in terms of 
the {\it primitive variables} $\{\rho, \epsilon, v_i \}$
(with $v_i$ the components of the velocity measured by 
an Eulerian observer) , as well as expressions for the 
source terms, are given in Banyuls {\it et al.} (\cite{BFIMM97}). 

  Let us consider the integral form of the above equations on a 
four--dimensional volume $\Omega \subset {\cal M}$ with three--dimensional 
boundary ${\partial \Omega}$, and apply Gauss theorem to obtain the 
corresponding balance equation

\begin{equation}
\int_{\partial \Omega} {\bf A} \cdot  d^3{\bf \Sigma} 
= \int_{\Omega} s d\Omega.
\label{intdiva}
\end{equation}

For numerical applications, we choose the volume $\Omega$ as the one bounded by
the coordinate hypersurfaces $\{\Sigma_{x^{\alpha}}$, $\Sigma_{x^{\alpha}+
\Delta x^{\alpha}}\}$. Hence, the time variation of the mean value of $A^0$,
\begin{equation}
\displaystyle{\overline A^0} = \frac{1}{\Delta {\cal V}}
\int^{x^1+\Delta x^1}_{x^1}
\int^{x^2+\Delta x^2}_{x^2} \int^{x^3+\Delta x^3}_{x^3}
\sqrt{-g} \, A^0 dx^1 dx^2 dx^3,
\end{equation}  
within the spatial volume 
\begin{equation}
{\Delta {\cal V}} = \int^{x^1+\Delta x^1}_{x^1} 
\int^{x^2+\Delta x^2}_{x^2} \int^{x^3+\Delta x^3}_{x^3} \sqrt{-g} 
dx^1 dx^2 dx^3 ,
\end{equation}
can be obtained from
\begin{eqnarray}
(\displaystyle{\overline A^0} {\Delta {\cal V}})_{t+\Delta t} = &
(\displaystyle{\overline A^0} {\Delta {\cal V}})_{t} + 
\displaystyle{\int_{\Omega}} s d\Omega \, - 
\nonumber \\ &
\left(
\displaystyle{\int_{\Sigma_{x^1}}} {\bf A} \cdot d^3{\bf \Sigma} +
\displaystyle{\int_{\Sigma_{x^1+\Delta x^1}}}{\bf A} \cdot d^3{\bf \Sigma} + 
\right.
\nonumber \\ &
\,\,\displaystyle{\int_{\Sigma_{x^2}}}{\bf A} \cdot d^3{\bf \Sigma} +
\displaystyle{\int_{\Sigma_{x^2+\Delta x^2}}}{\bf A} \cdot d^3{\bf \Sigma} + 
\nonumber \\ &
\left.
\,\displaystyle{\int_{\Sigma_{x^3}}}{\bf A} \cdot d^3{\bf \Sigma} +
\displaystyle{\int_{\Sigma_{x^3+\Delta x^3}}}{\bf A} \cdot d^3{\bf \Sigma} 
\right).
\label{integ}
\end{eqnarray}
\noindent

In order to update the solution in time, 
the volume and surface integrals on the
right hand side of Eq (\ref{integ}) have to be 
evaluated (see \S 3). HRSC schemes rely on the
calculation of the ${\bf A}$ vector fields by solving local 
Riemann problems combined with monotonized cell reconstruction 
techniques.

%%%%%%%%%%%%%%%%%%%%%%%%%%%%%%%%%%%%%%%%%%%%%%%%%%%%%%%%%%%%%

\section{Formulation of Riemann Problems in locally Minkowskian 
coordinates.}

According to the equivalence principle, physical laws in a 
{\it local inertial frame} of a curved spacetime have the 
same form as in special relativity.
Locally flat (or geodesic) systems of coordinates, 
in which the metric is brought into the Minkowskian form up to second order 
terms, are the realization of these local inertial frames. However, 
whereas the coordinate transformation leading to locally flat
coordinates involves second order terms, 
locally Minkowskian coordinates 
are obtained by a linear transformation.
This fact is of crucial importance when exploiting the selfsimilar 
character of the solution of the Riemann problems 
set up across the coordinate surfaces.
 
Hence, we propose to perform, at each numerical interface, 
a coordinate transformation 
to locally Minkowskian coordinates assuming 
that the solution of the Riemann problem is the one in special 
relativity and planar symmetry. This last assumption 
is equivalent to the approach followed in classical fluid 
dynamics, when using the solution of Riemann problems in slab symmetry for 
problems in cylindrical or spherical coordinates, which breaks down
near the singular points ({\it e.g.} the polar axis in cylindrical
coordinates). Analogously to classical fluid dynamics, 
the accuracy will depend on the magnitude of 
the Christoffel symbols, which might be large whenever
huge gradients or large temporal variations 
of the gravitational field are present. Finer grids and improved
time advancing methods will be required in those regions.

In the rest of this section we will focus on the evaluation of the 
first flux integral in Eq. (\ref{integ}).
\begin{eqnarray}
\label{intg1}
\int_{\Sigma_{x^1}} {\bf A}\cdot d^3{\bf \Sigma}=
\int_{\Sigma_{x^1}}{A^{1}}\sqrt{-g}\,dx^0dx^2dx^3 
\end{eqnarray}
To begin, we will express the integral on a 
basis ${\bf e}_{\hat{\alpha}}$ with
${\bf e}_{\hat{0}} \equiv {\bf n}$ and
${\bf e}_{\hat{i}}$ forming an orthonormal 
basis in the plane orthogonal to ${\bf n}$ with ${\bf e}_{\hat{1}}$
normal to the surface $\Sigma_{x^1}$ and
${\bf e}_{\hat{2}}$ and ${\bf e}_{\hat{3}}$ tangent to that surface.
The vectors of this basis verify
${\bf e}_{\hat{\alpha}}\cdot {\bf e}_{\hat{\beta}} = 
\eta_{\hat{\alpha}\hat{\beta}}$
with $\eta_{\hat{\alpha}\hat{\beta}}$ the Minkowski metric (in the
following, caret subscripts will refer to vector 
components in this basis).

Denoting by $x^\alpha_0$ the coordinates of the center of the interface
at time $t$, we introduce the following locally Minkowskian coordinate system
\begin{equation}
x^{\hat{\alpha}} = \left.M\right._\alpha^{\hat{\alpha}}(x^\alpha-x^\alpha_0), 
\label{change}
\end{equation}
where the matrix $M_\alpha^{\hat{\alpha}}$ is given by
$\partial_{\alpha}=M_{\alpha}^{\hat{\alpha}} {\bf e}_{\hat{\alpha}}$, 
calculated at $x^\alpha_0$.
In this system of coordinates the flux terms in the equations of GRH 
are written as in SRH, in Cartesian coordinates,
and the flux integral (\ref{intg1}) reads
\begin{eqnarray}
\label{intg2}
\int_{\Sigma_{x^1}}(A^{\hat{1}} - \frac{\beta^{\hat{1}}}
{\alpha}A^{\hat{0}})
\sqrt{-\hat{g}}\,dx^{\hat{0}}dx^{\hat{2}}dx^{\hat{3}} 
\end{eqnarray}
with $\sqrt{-\hat{g}} = 1+ {\cal O} (x^{\hat{\alpha}})$, 
where we have taken into account that, in the coordinates 
$x^{\hat{\alpha}}$, 
$\Sigma_{x^1}$ is described by the equation $x^{\hat{1}} - 
\frac{\beta^{\hat{1}}}
{\alpha} x^{\hat{0}} = 0$ (with $\beta^{\hat{i}} = M^{\hat{i}}_i \beta^i$),
where the metric elements ${\beta^{1}}$ and ${\alpha}$ are calculated 
at $x^{\alpha}_0$.
Therefore, the effect of a non-zero shift is that
the interface where the Riemann problem has to be solved
is not at rest but moves with {\it speed} 
$\,\,{\beta^{\hat{1}}}/{\alpha}$.

At this point, all the theoretical work
developed in the last years, concerning SRRS, can be exploited.  
The procedure involves the following steps:

1) We set up the Riemann problem by giving
the values at the two sides of $\Sigma_{x^1}$
of two independent thermodynamical 
variables (namely, the rest mass density $\rho$ and the specific internal 
energy $\epsilon$) and the components of the velocity in the orthonormal 
spatial basis $v^{\hat{i}}$, which are calculated using
\begin{equation}
v^{\hat{i}}=M_i^{\hat{i}} v^i
\end{equation}
where 
\begin{equation}
\displaystyle{
M_i^{\hat{i}} = \left( \matrix{
& \frac{1}{\sqrt{\gamma^{11}}}     
& \frac{-\gamma^{12} \gamma_{22}+\gamma^{13} \gamma_{23} }{\gamma^{11} \sqrt{\gamma_{22}}}     
& \frac{-\gamma^{13} \sqrt{\gamma_{22} \gamma_{33}-(\gamma_{23})^2} }{\gamma^{11} \sqrt{\gamma_{22}}}     
\cr
&   0   &  \sqrt{\gamma_{22}}    &  0   \cr
&   0   &  \frac{\gamma_{23}}{\sqrt{\gamma_{22}}}    
& \frac{\sqrt{\gamma_{22} \gamma_{33}-(\gamma_{23})^2} }{\sqrt{\gamma_{22}}}     
\cr
} \right) }
\end{equation}

2) The special relativistic 
Riemman problem is solved for the variables $\rho$, $\epsilon$
and $v^{\hat{i}}$, obtaining the fluxes associated to
${\bf J}, {\bf T}\cdot {\bf n}, {\bf T}\cdot {\bf e}_{\hat{j}}$. 
Notice that the effect of a non-zero
shift has to be considered at this stage.
Although most linearized Riemann solvers 
provide the numerical fluxes 
for surfaces at rest, it is easy to generalize them to moving 
surfaces relying on the conservative and hyperbolic character 
of the system of equations (see, {\it e. g.}, Harten and Hyman \cite{HH83}).

3) Once the Riemann problem has been solved,
by means of any linearized or exact SRRS, we can take advantage
of the selfsimilar character of the solution of the
Riemann problem, which makes it constant
on the surface $\Sigma_{x^1}$ simplifying
enormously the calculation of the above integral (\ref{intg2}):

\begin{equation}
\label{intg3}
\int_{\Sigma_{x^1}} {\bf A}\cdot d^3{\bf \Sigma}=
(A^{\hat{1}} - \frac{\beta^{\hat{1}}}
{\alpha}A^{\hat{0}})^{*}\int_{\Sigma_{x^1}}
\sqrt{-\hat{g}}\,dx^{\hat{0}}dx^{\hat{2}}dx^{\hat{3}} 
\end{equation}
\noindent
where the superscript (*) stands for the value 
on $\Sigma_{x^1}$ obtained from the solution
of the Riemann problem.
The quantity in parenthesis in Eq. (\ref{intg3}) represents 
the numerical flux across $\Sigma_{x^1}$.
Notice that the numerical fluxes correspond
to the vector fields 
${\bf J}$, ${\bf T}\cdot {\bf n}$, 
${\bf T}\cdot e_{\hat {1}}$, ${\bf T}\cdot e_{\hat {2}}$, ${\bf T}
\cdot e_{\hat {3}}$.
In order to obtain
the momentum fluxes in the original coordinates 
(${\bf T}\cdot \partial_i$)
the additional relation 
\begin{equation}
{\bf T}\cdot \partial_i=M_{i}^{\hat{j}}({\bf T}\cdot {\bf e}_{\hat{j}}) 
\end{equation}
has to be used.

4) Finally, the numerical fluxes are multiplied by the surface integral
appearing at the right hand side of (\ref{intg3}), that is
expressed in terms of the original coordinates as
\begin{equation}
\label{intg4}
\int_{\Sigma_{x^1}}\sqrt{\gamma^{11}}
\sqrt{-{g}}\,dx^{{0}}dx^{{2}}dx^{{3}} 
\end{equation}
and can be easily evaluated for a given metric.

\bigskip
 
Let us remind that, in this section, we have focussed on the 
calculation of the flux terms in Eq. (\ref{integ}), for given
left and right states. Obviously, the performance of the numerical
code depends on the quality of the provided initial states, as well
as the computation of the source terms in Eq. (\ref{integ}), and
the algorithm for time advancing. In all the calculations presented
in next section, left and right states for Riemann problems have been 
obtained with a monotonic, piecewise linear reconstruction procedure.
The source integrals have been evaluated assuming constant values
of $\rho$, $\epsilon$ and $v^i$ inside the numerical cells. Finally,
advance in time has been done by means of a TVD-preserving, third
order Runge-Kutta, that takes into account the influence of the
source terms in the intermediate steps. 
Notice that the treatment of the source terms
is essential for the method to work in regions where they dominate.
A treatment consistent with the reconstruction procedure and 
better time advancing schemes may be required in regions very close
to coordinate singularities, where the source terms might diverge.
 
%%%%%%%%%%%%%%%%%%%%%%%%%%%%%%%%%%%%%%%%%%%%%%%%%%%%%%%%%%%%%
\section{Tests and applications.}
In order to demonstrate the feasibility of our procedure 
we have considered an exhaustive 
sample of standard discontinuous initial value problems for which 
the exact solution is known, as well as some numerical applications 
involving strong gravitational fields. 
The set of SRRS used in the computations are
the exact one (Mart\'{\i} and M\"uller \cite{MM94}) for 1D problems, 
as well as SR extensions of the linearized solvers described in
Harten, Lax and van Leer (\cite{HLLE}), Roe (\cite{R81}),
and Donat and Marquina (\cite{DM96}). 

To summarize,
we have successfully redone all the 
experiments shown in Banyuls {\it et al.} (\cite{BFIMM97}), 
that includes relativistic shock-tube tests for 
non-diagonal metrics, as well as 
a number of simulations of relativistic
wind accretion onto a Schwarzschild black hole.
In Figure 1, we show the results from a simulation of spherical
accretion of an ideal gas onto a Schwarzschild black hole. The 
analytical solution derived by Michel (\cite{Mic72}) is represented
by the solid line and the diamonds represent the numerical solution
obtained using the exact SRRS, after the stationary state has
been reached. In Figure 2, we display results from two dimensional
simulations of non-spherical accretion onto a moving black hole, 
corresponding to one of the models recently studied in 
Font \& Iba\~{n}ez (\cite{FI98}) [model MC2 in their table I].
The figure displays isocontours of the rest mass density in logarithmic
scale. The upper-left panel displays the results obtained from the
code described in Banyuls {\it et al.} (\cite{BFIMM97}), the 
upper-right panel shows the results obtained with the new approach
using the same solver (Roe-like). Results obtained with the
new approach using HLLE and Marquina's solvers are shown in the
lower-left and lower-right panels, respectively.
              
The main conclusion emerging from the comparison is that
our new approach generates remarkably similar results for 
the four different SRRS, 
the tiny differences being due to the intrinsic
properties of each solver, e.g., Roe's
solver is the least diffusive and therefore more oscillatory. 
The results following the method presented in 
Banyuls {\it et al.} (\cite{BFIMM97}) are
indistinguishable from the ones obtained
using the special relativistic Roe-like solver in this new approach.
It can be shown analytically that
both algorithms are equivalent.

%%%%%%%%%%%%%%%%%%%%%%%%%%%%%%%%%%%%%%%%%%%%%%%%%%%%%%%
\section{Conclusions and outlook}
We have developed a general procedure  
to use SRRS in multidimensional general relativistic 
hydrodynamics that allows us to take 
advantage of the increasing number of SRRS developed recently,
overcoming partial approaches derived in previous
papers, which were restricted to linearized Riemann solvers. 
Since the change of coordinates we propose is linear 
and only involves a few arithmetical operations,
the additional computational cost of the approach is completely 
negligible.

The procedure has a large potentiality and could be applied to 
other systems of conservation laws, as the equations of
general relativistic magneto-hydrodynamics,
providing a very useful numerical tool to solve them
using the corresponding Riemann solvers 
developed for the special relativistic case.
The feasibility of the approach has been extensively tested and
its numerical performance is, at least, as good as other schemes
developed in previous papers, having the additional advantage of being
very well suited to include future work and improvements
that might be done in the field of SR Riemann solvers. 

%%%%%%%%%%%% 

\acknowledgements{ This work has been supported by the Spanish 
DGICYT (grant PB94-0973).
J. M$^{\underline{\mbox{a}}}$. M. and J.A.F have benefited from 
European  contracts (nrs. ERBFMBICT950379 and ERBFMBICT971902, respectively) .
J.A.F. also acknowledges previous financial support from the 
MPG to work at the AEI in Potsdam. 
The authors thank M.A. Aloy, S. Brandt, J. Ferrando, P. Papadopoulos,
M. Portilla and E. Seidel for useful discussions. }

%%%%%%%%%%%% 

{}

\newpage

%\begin{figure}
{\bf Fig.1.}{ Spherical accretion of an ideal gas: profiles of density 
($\rho$), internal energy ($\epsilon$) and velocity ($v/c$)
as a function of the radial
coordinate, after the steady state has been reached. 
The critical point ($r_c$), the critical density at the critical
point ($\rho_c$) and the adiabatic exponent of the equation of state ($\gamma$)
have been taken $r_c=200 M$, $\rho_c = 7 \times 10^{-4}$ and $\gamma=4/3$.
The solid line
corresponds to the analytic solution and the diamonds to the numerical
one obtained using the exact SRRS.}
%\end{figure}

\bigskip

%\begin{figure}

{\bf Fig.2.}
{Non-spherical hydrodynamic accretion onto a Schwarzshild
black hole. The initial model is characterized by an asymptotic
velocity of $0.5c$ and Mach number $5$. The adiabatic exponent of
the fluid is $5/3$. The simulation employs a grid of $120 \times 40$
zones in the radial and polar directions respectively. The radial
domain extends from $2.1M$ to $38M$. The polar domain extends
from $0$ to $\pi$. The flow moves from left to right. The different
pannels show isocontours of the logarithm of the density normalized
to its asymptotic value. Starting from the upper-left pannel and in a
clockwise sense we show results for the Roe solver (as used in
Banyuls et al (1997)), its SRRS version, Marquina's solver and HLLE.
The isocontours span the same interval regardless of the solver used.
This range goes from $0$ to $1.15$. The maximum values are always
found at the rear part of the hole where the matter piles up. The
different solvers agree on this value up to three significant figures.}
%\end{figure}

\newpage

\begin{figure}
\psfig{figure=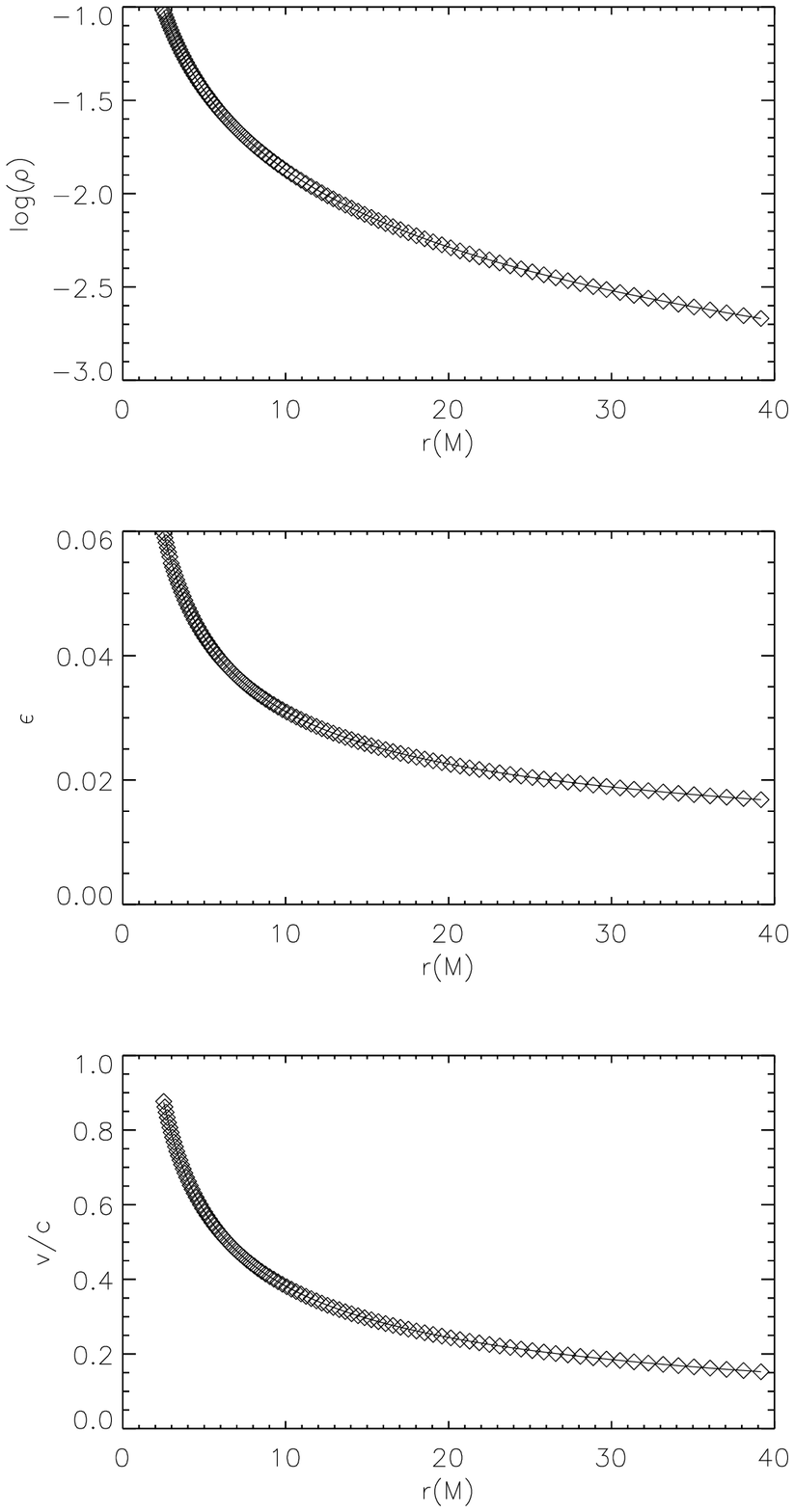}
\end{figure}

\newpage

\begin{figure}
\psfig{figure=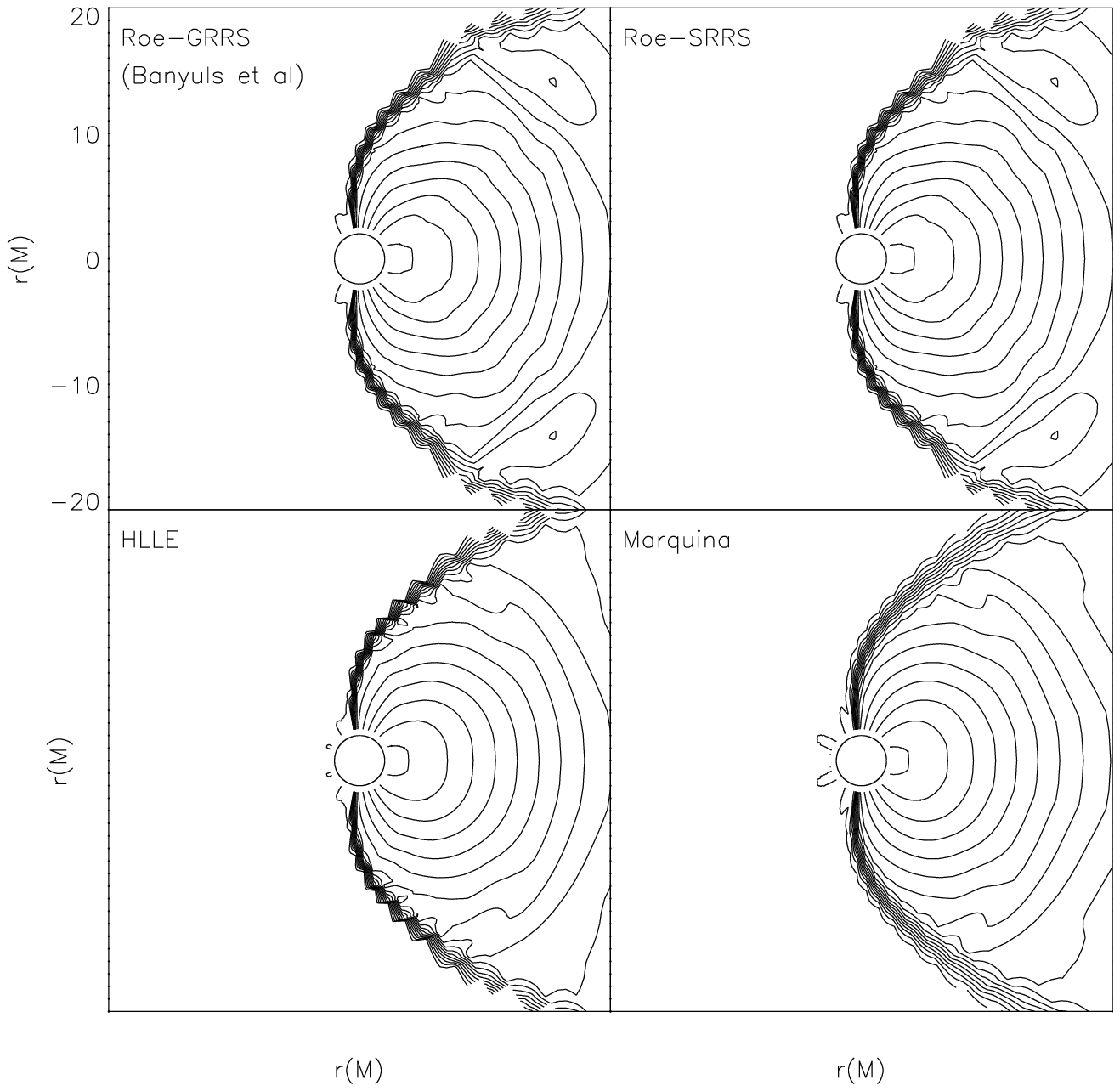}
\end{figure}


\begin{thebibliography}{}


\bibitem[1995]{Abr95} Abrahams, A.,  Anderson, A., 
Choquet-Bruhat, Y., and York, J.W., 1995, 
Phys. Rev. Lett.,75, 3377.

\bibitem[1997]{BFIMM97} Banyuls, F., Font, J. A., 
Ib\'a\~{n}ez, J. M$^{\underline{\mbox{a}}}$., 
Mart\'{\i}, J. M$^{\underline{\mbox{a}}}$., and Miralles, J.A.,
1997, ApJ, 476, 221.

\bibitem[1995]{Bon95} Bona, C., Mass\'o, J., Seidel, E., 
and Stela, J., 1995, Phys. Rev. Lett., 75, 600. 

\bibitem[1993]{escorial} Bona, C., 
Ib\'a\~{n}ez, J.M$^{\underline{\mbox{a}}}$.,
Mart\'{\i}, J.M$^{\underline{\mbox{a}}}$., and Mass\'o, J., 1993, 
in {\it Gravitation and General Relativity: Rotating Bodies and Other 
Topics}, edited by F. Chinea and L.M. Gonz\'ales-Romero
(Springer Verlag, Berlin).

\bibitem[1996]{DM96} Donat, R. and Marquina, A., 1996, 
J. Comput. Phys., 125, 42.

\bibitem[1994]{EM94} Eulderink, F., and Mellema, G., 1994, 
A\&A, 284, 652.


\bibitem[1994]{FIMM94} Font, J.A., 
Ib\'a\~nez, J. M$^{\underline{\mbox{a}}}$., 
Marquina, A. and Mart\'{\i}, J.M$^{\underline{\mbox{a}}}$., 1994,
A\&A, 282, 304.

\bibitem[1998]{FI98} Font, J.A.
and Ib\'a\~nez, J. M$^{\underline{\mbox{a}}}$., 1998,
ApJ, 494, 297.


\bibitem[1985]{Fri85} Friedrich, H., 1985, 
Comm. Math. Phys., 100, 525.

\bibitem[1996]{Fri96} Fritelli, S., and  Reula, O., 1996,
Phys. Rev. Lett., 76, 4667.

\bibitem[1959]{Go59} Godunov, S.K., 1959, Mat. Sb.,  47, 271.

\bibitem[1983]{HH83} Harten, A. and Hyman, J.M., 1983,  
J. Comput. Phys., 50, 235.

\bibitem[1983]{HLLE}
Harten, A., Lax, P.D., and van Leer, B., 1983, SIAM Review, 25, 35.

\bibitem[1997]{IFMM97} 
Ib\'a\~nez, J. M$^{\underline{\mbox{a}}}$., {\it et al.}, 1997,
in {\it Proceedings from the 18th. Texas Sym. on Relativistic Astrophysics},
World Scientific Press.

\bibitem[1991]{MIM91} 
Mart\'{\i}, J.M$^{\underline{\mbox{a}}}$.,
Ib\'a\~nez, J. M$^{\underline{\mbox{a}}}$., and Miralles, J.A., 1991, 
Phys. Rev., D43, 3794.

\bibitem[1994]{MM94} 
Mart\'{\i}, J.M$^{\underline{\mbox{a}}}$.,
and M\"uller, E., 1994, J. Fluid Mech., 258, 317.

\bibitem[1994]{MMI94} 
Mart\'{\i}, J.M$^{\underline{\mbox{a}}}$.,
M\"uller, E., and Ib\'a\~nez, J. M$^{\underline{\mbox{a}}}$., 1994, 
A\&A, 281, L9.

\bibitem[1972]{Mic72} 
Michel, F. C., 1972, Ap\&SS, 15, 153.

\bibitem[1986]{NW86} Norman, M.L., and Winkler, K-H.A., 1986,
{\it Astrophysical Radiation Hydrodynamics}, (Reidel).

\bibitem[1996]{Retal96} Romero, J.V., 
Ib\'a\~nez, J. M$^{\underline{\mbox{a}}}$.,
Mart\'{\i}, J.M$^{\underline{\mbox{a}}}$., and Miralles, J.A., 1996,
ApJ, 462, 836.

\bibitem[1981]{R81} Roe, P.L., 1981, J. Comput. Phys., 53, 357.

\bibitem[1979]{Wi79} Wilson, J.R., 1979, in {\it Sources of gravitational
radiation \rm}, (Cambridge University Press).

\end{thebibliography}
\end{document}